 \documentclass[final,5p,times,twocolumn]{elsarticle}

 \usepackage{graphicx}

\usepackage{amssymb}

 \usepackage{lineno}

 \biboptions{sort&compress}

\journal{Astroparticle Physics}

\begin{document}

\begin{frontmatter}



\title{Additional experimental evidence for a solar influence on nuclear decay rates}


 \author[PurdueNE,PurduePhys]{Jere H. Jenkins\corref{cor1}}
 \address[PurdueNE]{School of Nuclear Engineering, Purdue University, 400 Central Dr., West Lafayette, IN  47907 USA}
 \address[PurduePhys]{Department of Physics, Purdue University, West Lafayette, IN 47907}
 \ead{jere@purdue.edu}
 \cortext[cor1]{Corresponding author}
 \author[OSURR]{Kevin R. Herminghuysen}
 \address[OSURR]{Ohio State University Research Reactor, The Ohio State University, Columbus, OH 43210 USA}
 \author[OSURR]{Thomas E. Blue}
 \author[PurduePhys]{Ephraim Fischbach}
 \author[Animal]{Daniel Javorsek II}
 \address[Animal]{412th Test Wing, Edwards AFB, CA 93524, USA}
 \author[OSURR]{Andrew C. Kauffman}
 \author[Mayo]{Daniel W. Mundy}
 \address[Mayo]{Department of Radiation Oncology Physics, Mayo Clinic, Rochester, MN 55905 USA}
 \author[Stanford]{Peter A. Sturrock}
 \address[Stanford]{Center for Space Science and Astrophysics, Stanford University, Stanford, CA 94305 USA}
 \author[OSURR]{Joseph W. Talnagi}
 
\begin{abstract}
Additional experimental evidence is presented in support of the recent hypothesis that a possible solar influence could explain fluctuations observed in the measured decay rates of some isotopes. These data were obtained during routine weekly calibrations of an instrument used for radiological safety at The Ohio State University Research Reactor using $^{36}$Cl. The detector system used was based on a Geiger-M\"{u}ller gas detector, which is a robust detector system with very low susceptibility to environmental changes. A clear annual variation is evident in the data, with a maximum relative count rate observed in January/February, and a minimum relative count rate observed in July/August, for seven successive years from July 2005 to June 2011. This annual variation is not likely to have arisen from changes in the detector surroundings, as we show here.
\end{abstract}


\end{frontmatter}



\section{Introduction}
\label{Intro}

Evidence for a possible solar influence on nuclear decay rates has recently been presented based on the analysis of decay rate measurements taken at three independent institutions. The first was an apparent change in the measured decay rate of $^{54}$Mn during a series of solar flares in December of 2006 \cite{jen09a}. The $^{54}$Mn data were being collected as part of a half-life measurement utilizing continuous four-hour measurements. This allowed a time resolution capable of seeing changes that could have been caused by a solar flare, which typically lasts minutes to hours. This work was then followed by two additional papers by our group \cite{jen09b,fis09}, where data were analyzed from half-life measurements taken by two independent groups, one at the Brookhaven National Laboratory (BNL) in Upton, New York, USA, and the other at the Physikalisch-Technische Bundesanstalt (PTB) in Braunschweig, Germany. The BNL group had undertaken a measurement of the half-life of $^{32}$Si, and the data from that experiment exhibited  a periodic oscillation with an approximate period of 1 year \cite{alb86}. The measurements taken at the PTB in Germany were of a $^{226}$Ra standard used for comparison in the measurements of the half-life of $^{152}$Eu \cite{sie98}, and in a longer-term analysis of the stability of detectors used in standards laboratories. The $^{226}$Ra data also showed a periodic oscillation, again with a period of approximately 1 year. 

In the subsequent analysis of the raw data obtained from the BNL and PTB experiments, both data sets were shown to have not only the same period, but in the two years during which the data sets overlapped they had the same approximate phase and amplitude as well \cite{jen09b,jav10}. Moreover, it was shown that both data sets were not only in phase with each other, but also appeared to be approximately in phase with the distance of the Earth from the Sun. Taking all of these experiments into account (BNL, PTB and Purdue), a reasonable case could be constructed for the possibility of a solar influence on nuclear decays \cite{fis09}. This case was subsequently strengthened as a result of an analysis by our group \cite{stu10a,stu10b}, where an additional periodicity was identified in the BNL data at 11.25$\pm$0.07/yr, and in the PTB data at 11.21$\pm$0.13/yr. Both of these peaks may be linked to the rotation (and probable inhomogeneous nature) of the core of the Sun \cite{stu99,stu08,stu09}. A third periodicity was also identified by our group \cite{stu11a} in both the BNL and PTB data sets, which is analogous to the Rieger periodicity \cite{rie84} with a period of approximately 173 days. An analysis of the phases of these periodicities was also carried out \cite{stu11b} which determined that the phase characteristics of the annual periodicities could reasonably be attributed to a solar influence on the decay rates.

The suggestion of a solar influence on nuclear decay rates has been met with some criticism, however. An analysis 
by \citet{nor09} of decay data taken for several isotopes in their laboratory did not see evidence for an annual effect similar to that reported in Ref. \cite{jen09b}. \citet{sil09} examined data from $^{22}$Na decay measurements, and also found no similar fluctuations. \citet{coo09} analyzed the heat output data from the radioisotope
thermal generators (RTGs) onboard the Cassini spacecraft, and found no evidence for time
variation in the decay of $^{238}$Pu. Based upon the absence of oscillations
similar to those found in the BNL and PTB datasets, both Norman, et al., and Cooper
concluded that the solar influence suggested in Refs. \cite{jen09a,jen09b,fis09} was not present. However, a more recent
analysis \cite{oke11} of the data of
\citet{nor09} does suggest the presence of a solar influence, albeit at a
lower level than indicated by the BNL or PTB data. As has been noted
previously \cite{jen10}, the very same 
nuclear physics considerations which are
responsible for the fact that beta-decay half-lives vary from fractions of a 
second to billions of years (i.e., nuclear wavefunctions, selection rules, 
phase space, etc.) would also apply to the effects of any solar influence. Hence, 
there should be no expectation that periodic effects would be present in all 
beta-decays at the same $\sim3\times10^{-3}$ level seen in the BNL and PTB data. 
This observation applies as well to the data in Ref. \cite{sil09}, and 
especially to the analysis in Ref. \cite{coo09}. In the latter case, the decay 
of $^{238}$Pu is a pure alpha-decay, and leads to a daughter ($^{234}$U) which is an alpha decay 
with a 246,000 year half-life \cite{nndc}. Therefore, there would have been no significant 
contributions from beta-decays in the Cassini RTG data, whereas all previous data sets 
in which periodic effects were seen were measurements of beta-decays or electron capture.

Similar anomalous behaviors and periodicities in nuclear decay data have in fact been observed by other groups. Data from the measurement of $^{60}$Co and $^{90}$Sr/$^{90}$Y published by Parkhomov \cite{par05,par10a,par10b} also exhibit annual and monthly periodicities when measured on separate Geiger-M\"{u}ller (G-M) counting systems in a controlled experiment. Interestingly, $^{239}$Pu counted by Parkhomov did not show any such periodicities. This demonstrates two important points: first, that the oscillations were likely not of an environmental origin; and second, the oscillations appear to arise primarily in beta-decays, in agreement with the previous remarks. Also, since the counting systems were G-M detectors in the Parkhomov experiments, there would be no environmentally induced gain shifting since there is no amplifier in the system. 

Further evidence of annual periodicities in decay data was presented in an earlier publication by Ellis \cite{ell90}. His data showed an annual oscillation in the measured decay rate of neutron-activated manganese foils used to calibrate a system of plutonium-beryllium neutron sources. Interestingly, the $^{56}$Mn counts exhibited an annual periodicity, yet the $^{137}$Cs standard used to calibrate his scintillation detection system did not. This indicates that, as an experimental observation, isotopes have different sensitivities to whatever influence is causing the observed effects. Moreover, since the two isotopes were measured on the same counting system, it would also appear to rule out a simple environmental systematic cause. This supports the analysis by Jenkins, Mundy and Fischbach \cite{jen10}, who examined all of the likely environmental influences on the counting systems utilized in the BNL, PTB and Purdue experiments \cite{jen09b,alb86,sie98} and concluded that all of the known suspect effects were too small to have caused the observed oscillatory behavior. Additionally, a recent paper by \citet{ste11} presents extensive evidence for annual and sub-annual periodicities in the measured decay rates of $^{222}$Rn (and its progeny), which lends further support to the solar influence hypothesis. Furthermore, we performed additional analyses \cite{stu12} of the $^{90}$Sr/$^{90}$Y data published in Refs. \cite{par05,par10a,par10b}, and found that the data which contained annual and monthly periodicities also exhibited striking similarities to the frequency content in the Mount Wilson Solar Observatory's solar diameter measurement data.

Some recent experiments have been performed to test the hypothesis of a neutrino-mediated solar influence on terrestrial nuclear decays by two independent groups, two of them performed by our group and one performed by an independent group in South Africa. The two experiments at the National Institute of Standards and Technology (NIST) performed by our group \cite{lin10,lin11} examined a variation of the electron anti-neutrino flux ($\bar{\nu}_e$) resulting from the $\beta^{-}$-decay of $^{198}$Au by utilizing samples with different geometries.  The theory behind this experiment is that if the specific activity was high enough from the resulting decay of $^{198}$Au in a sphere (compared to a foil as in Ref. \cite{lin10} or a wire as in Ref. \cite{lin11}), the $\bar{\nu}_e$ flux could approach the solar electron neutrino (${\nu}_e$) flux experienced on Earth, which is $\sim60\times{}10^{9}~\nu_e ~cm^{-2}s^{-1}$. Although the results of both of these experiments were inconclusive, they did not clearly support the hypothesis that $\bar{\nu}_e$ could affect  the $^{198}$Au $\beta^{-}$-decay. A different experiment was performed by \citet{dem11}, where decay rates of various isotopes were measured in close proximity to a 2 MW$_{th}$ nuclear reactor, which is a well characterized sources of $\bar{\nu}_e$. The negative results of this series of experiments led \citet{dem11} to suggest two possibilities. The first was that the $\bar{\nu}_e$ flux was not high enough from the 2 MW$_{th}$ reactor, and a larger reactor may show better results. The second, which was similar to the conclusions drawn by our group in Refs. \cite{lin10,lin11}, was that $\bar{\nu}_e$ might not have the same effect as $\nu_e$ on $\beta^{-}$-decay. Furthermore, the possibility exists that solar $\nu_e$ are not primarily responsible for the observed effect, but rather some other component of solar neutrino flux (e.g., $\nu_{\mu}$ or $\nu_{\tau}$) or an as yet unknown particle or field \cite{lin11}.

\begin{figure*}[ht]
\includegraphics[scale=0.70]{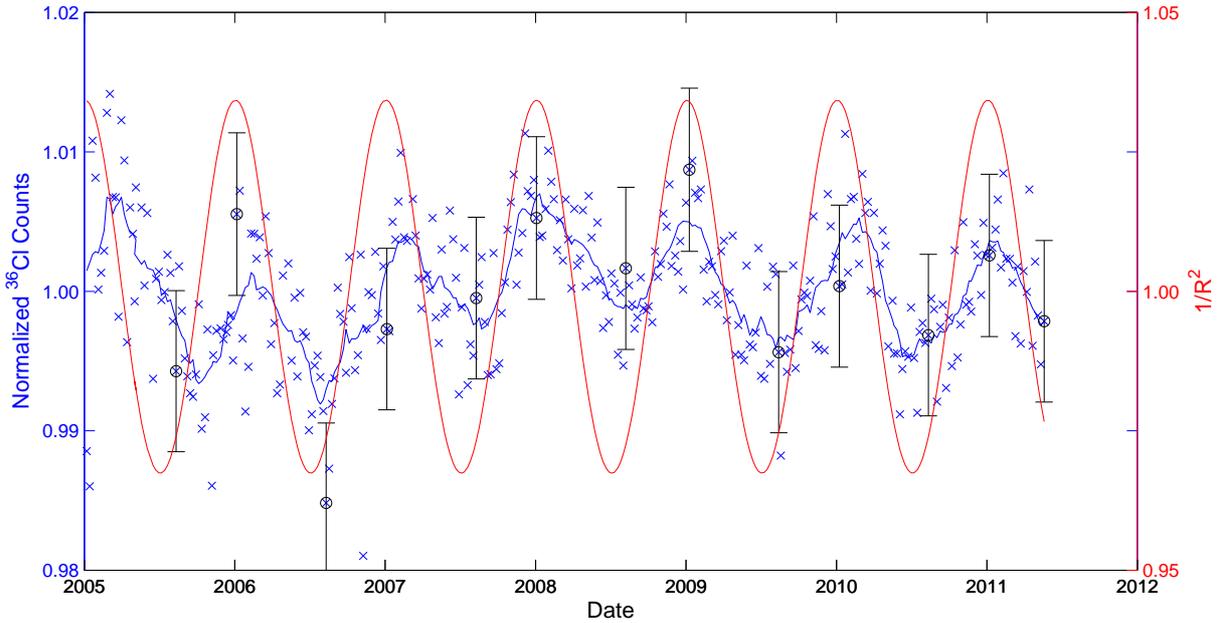}
\caption{Plot of the measured $^{36}$Cl decays taken at the Ohio State University Research Reactor (OSURR). The crosses are the individual data points, and the blue line is an 11-point rolling average centered on each point. The red curve is the inverse of the square of the Earth-Sun distance. Error bars are shown for a limited number of $^{36}$Cl points in order to maintain readability. \label{fig:data}}
\end{figure*}

The purpose of this article is to present $^{36}$Cl decay data collected at The Ohio State University Research Reactor (OSURR), in Columbus, Ohio, USA, over the course of 7 years, which further strengthen the case for a solar influence on some nuclear decays. The data were taken weekly, as part of the calibration check of an instrument used at the OSURR facility, thus the data were not the result of an experiment per se, but were collected as part of routine operations at the OSURR. It is evident from the data shown in Fig.~\ref{fig:data}  that there is an oscillation with an approximate annual period that appears to correlate with the inverse-square of the Earth-Sun distance, and possibly additional frequencies that can also be linked to the Sun.

\section{Data Collection/Detector Set-up}
\label{Data}

As noted above, the $^{36}$Cl data were collected weekly as part of the efficiency check of the Eberline Beta Counter BC-4 that is used for counting of contamination survey wipes. The detector system incorporates a 1.75 inch (4.4 cm) diameter G-M pancake-style tube contained inside 0.875 inches (2.22 cm) of lead to reduce background \cite{ebe72}. The BC-4 instrument itself sits in the reactor bay at OSURR, which has partial environmental control.  Although it is not air-conditioned, space heaters are used to maintain the interior temperature in a comfortable range during colder months. Thus, temperature is controlled to some extent, but relative humidity is not. 

The source utilized for the calibration check is a 0.4 $\mu$Ci $^{36}$Cl split check source (manufactured by Nuclear-Chicago, Model SK2-1) with a diameter of 1.0 inch (2.54 cm). The two aluminum half-disk sources which comprise the check source are contained within an aluminum holder with an outer diameter of 1.25 inches (3.2 cm).  The active regions of the source are two machined circular depressions, each approximately 0.2 inches in diameter, located near the center of the 1-inch disk on each half disk. The decay of $^{36}$Cl is primarily by $\beta^{-}$-emission (98.1\%, $E_o$=708.6(3) keV, $T_{1/2}=3.01(2)\times10^5$ y \cite{end99}) to the ground state of $^{36}$Ar, which is stable (there is also a competing K-capture mode to $^{36}$S with a 1.90\% branching ratio). We note in passing that this is the same isotope used as the standard in the BNL $^{32}$Si half-life experiment.

After early 2006, the source geometry within the BC-4 counter was controlled by an aluminum disk insert with an outer diameter of approximately 2.0 inches. This disk centered the source within the 2-inch planchet used to place the source on the detector tray, and this ensured that the check source was located in the same position under the detector for each calibration-check measurement. The source/insert combination was then placed into the planchet, and then the planchet was placed into the source tray in the BC-4, which slides in to locate the source under the detector within the counter. Geometry in the counter is well controlled, such that the source-detector separation is very small ($\sim$2 mm). The slide tray position is controlled by a stop at the rear of the detector system, thus controlling the y-axis geometry.

Data recorded each week were the gross counts (source with background, with no dead-time correction), and a ten minute background count. Typical $^{36}$Cl count rates were approximately 124,600/120s ($\sim$1038 counts/s), with a fractional uncertainty of $\sim$0.28\%. Typical background count rates were approximately  225 counts/600s ($\sim$0.35 counts/s), with a fractional uncertainty of $\sim$6.7\%. Since the half-life of $^{36}$Cl is much longer than the duration of the data series reported here, the standard decay rate can be assumed to have been approximately constant during the course of the experiment.

\section{Results/Discussion}

The measured $^{36}$Cl data are shown in Fig.~\ref{fig:data}. These data points are the weekly counts from 7 January 2005 to 17 June 2011, a total of 334 points. Since the $^{36}$Cl decay rate can be assumed to be relatively constant over the five years of the series reported here, the only adjustment to the data presented in Fig.~\ref{fig:data} is to normalize each measured count by dividing the measured counts by the average of all 334 counts ($\bar{x}=124593, \sigma=632$). Note that the consistency of the data improves after early 2006 when the source geometry control was added. An 11 point moving average of the data is has been added to the figure as a solid blue curve to aid in the visual presentation.

It is clear from Fig.~\ref{fig:data} that the data exhibit an annual periodicity, with higher counts in winter (approximately February), and lower counts in the summer (approximately August). This periodicity is very similar to the annual oscillations presented in the $^{32}$Si/$^{36}$Cl data from BNL \cite{alb86,jen09b} and the $^{226}$Ra and progeny data from the PTB \cite{sie98,jen09b}. Further analysis reveals that the normalized counts shown in Fig.~\ref{fig:data} oscillate with an approximate annual period that is roughly correlated with the annual period of the inverse-square of the Earth-Sun distance, which is shown as the red curve in Fig.~\ref{fig:data}.

\begin{figure}[h]
\includegraphics[width=\columnwidth]{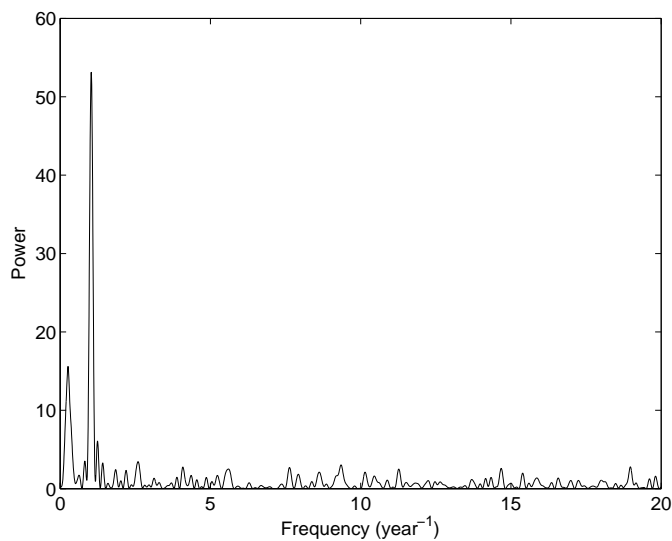}
\caption{Frequency spectrum for the OSURR $^{36}$Cl data. The annual oscillation is clearly evident, as is a peak at $\sim$0.25/yr that is probably related to the length of the data set. None of the other peaks are statistically significant. \label{fig:spec}}
\end{figure}

To verify the presence of the annual periodicity, a power-spectrum analysis was performed for the data in Fig.~\ref{fig:data}. In power-spectrum analysis, the probability of the null hypothesis 
(that the peak is due to normally distributed random noise) is given by $e^{-S}$ (where $S$ is the power value of the frequency peak)  \cite{sca82}. These results are presented in Fig.~\ref{fig:spec}, with frequency in units of inverse years.  The prominent peak in the figure for a frequency of 1 yr$^{-1}$ (with a power of $\sim$54) confirms the annual periodicity.  The peak in the power spectrum at $\sim$0.25 yr$^{-1}$ is probably related to the length of the data set, but may represent the influence of the $\sim$11-year solar cycle.  None of the other peaks in the power spectrum are statistically significant.
Although the annual periodicity shown in Fig.~\ref{fig:spec} is quite prominent, there remains a question as to the cause of this annual oscillation. Is the annual periodicity related to some property of the Earth-Sun system, such as the inverse-square of the Earth-Sun distance, or rather is it a consequence of interfering or modifying inputs to the counting system that vary periodically with a frequency of 1 yr$^{-1}$? Background radiation is an example of an interfering input, and the dependence of the detector, counting system, and the transport of radiation across the source/detector gap on environmental factors (such as temperature and humidity) is an example of a modifying input. 

To address these concerns, we begin by analyzing the detector system utilized in this experiment, which is G-M tube as described previously. A G-M tube is a gas detector with an inert gas (usually argon or neon) quenched with a halogen gas in this case. The use of a halogen quench over an organic compound as the quenching gas allows for a recombination of the diatomic halogen, and extends the life of the tube indefinitely. When radiation, such as an $\alpha$- or $\beta$-particle, or an energetic electron from a photon interaction in the detector wall, interacts within the tube gas, an electron-ion pair is generated. Photons can interact directly with the working gas and cause ionization, but this is a lower probability interaction. The bias of the detector provides a driving force for the electrons to move to the anode, and the ions to move to the cathode. As an electron is accelerated by the very strong electric field within the detector, it can accelerate to the point where it generates additional ionizations, which result in an avalanche of ionizations within the working gas. The avalanche of ionizations continues to the point where it self-terminates due to a reduction of the electric field due to the presence of so many positively charged ions. As a result, essentially all of the gas that can be ionized is ionized with each discharge.

The large number of electrons collected at each discharge provide a charge pulse of sufficient magnitude that it does not require amplification to be counted, and therefore there are no pre-amplifiers or amplifiers required in the processing of the signal out of the detector. A trigger circuit is the next step in the system, and as long as the pulse out of the detector is of large enough magnitude, a standard logic pulse is output, which is counted by a scaler-timer. While the disadvantages of the G-M detector are mostly along the lines of its inability to perform energy-spectroscopy or discrimination between radiation types, the significant advantage is that it is a robust, very stable system for detecting radiation events. Therefore it is suitable for relatively harsh working environments, and is a very practical, stable system. The general stability of G-M systems, similar to the one used to collect these data, with respect to temperature has been well studied \cite{fuj51,haq83,khr60,kim50,kim51,lie48,sei52a}. These analyses determined that the stability of the G-M system was not likely to be a factor within the range of temperatures that were likely to be experienced in the OSURR laboratory (65-85$^{\circ}$F, though the actual temperature range experienced was likely much smaller). We can thus conclude that a change in the detector system is not a likely candidate for the source of the fluctuations.

\begin{figure}[h]
\includegraphics[width=\columnwidth]{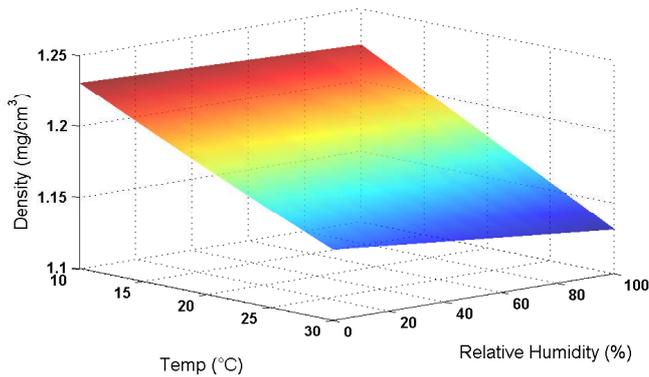}
\caption{Plot of air density as a function of temperature and relative humidity as adopted by CIPM \cite{dav92,pic08}. \label{fig:airden}}
\end{figure}

In light of the fact that beta-particles lose energy with each interaction with an air molecule, the transport of the beta-particles across the source-detector gap clearly depends on the density of the intervening air. This very argument, that air density was a factor in changing the measured count rates, was one of the bases proposed by \citet{sem09} to question the claims in Ref. \cite{jen09b}. \citet{sem09} assumed that humid air, such as would be experienced in the summer months, was denser than dry, winter air. This is not the case, as shown in Fig.~\ref{fig:airden}, which is a plot of air density as a function of temperature and moisture content (measured by relative humidity) as adopted in CIPM-2007 \cite{dav92,pic08}. As can be seen from the figure, cooler, drier air is significantly denser than warm, humid air. Therefore, based on air density, if the count rates were to be affected by the energy loss of the beta particles due to denser air, one could assume that the count rates should be \textit{lower} in the winter when the inside air is coolest and driest. Examining Figure \ref{fig:data}, this is clearly not the case. In fact, the counts are actually highest when the air is the coolest and driest (i.e., the highest density).

It is still worthwhile to examine the beta particle transport across the source-detector gap using a radiation transport code, such as MCNP (something that was suggested as well in Ref. \cite{sem09}), even though the higher count rates in the cooler, drier, denser winter air imply that beta particle transport is not an issue across the very small gap between source and detector. Such an MCNP analysis was performed for the BNL source-detector system as a part of a larger, detailed analysis of environmental influences by our group on the BNL and PTB detector systems in Ref. \cite{jen10}. The BNL source/detector system (a gas proportional detector) was modeled using the Monte Carlo N-Particle eXtended (MCNPX, \cite{mcnp}) code for both the $^{36}$Cl and $^{32}$Si samples. The $^{36}$Cl source-detector distance in the BNL experiments was 4.000 mm, which is almost twice the distance between the source and detector for the OSURR BC-4 (2 mm). The results for the $^{36}$Cl analysis from Ref. \cite{jen10} are presented in Table \ref{table1}. For each of the parameters analyzed, it can be seen from these results that any possible variation (measured in change per degree Fahrenheit, or $\Delta{}/^{\circ}\mathrm{F}$) of parameters (e.g., energy deposition in the detector volume, particle current, etc.) is at least two orders of magnitude too small to have caused the observed periodic effects. This supports the conclusions of \citet{alb86} in their own analysis of possible environmental and systematic influences on their detector system. In Table 1 of Ref. \cite{alb86}, limits were also set on other systematic influences at less than a 10$^{-4}$ fractional change for each parameter, which is too small to have explained the oscillations they observed. We can thus safely say, based on the above analysis, that air density changes were not likely to have affected the transport of particles across the 2 mm source-detector gap, nor should it have affected the deposition of energy by the beta particles in the detector volume. This leaves one additional parameter of a gas detector system, which is the energy needed to ionize the working gas. Since it is known that a G-M detector only requires a single ionization to discharge essentially the entire tube, and also that the ionization potential of argon (the working gas) does not change for pressures between 1 and 200 bars \cite{fou20}, we can again safely conclude that a G-M tube is insensitive to small changes in the ambient laboratory pressures.

\begin{table}

\label{table1}
\caption{Table of MCNPX results showing the sensitivities for various parameters of the BNL source-detector system, per degree Fahrenheit. \cite{jen10}}
\scriptsize
\begin{tabular}{ l c c }
  & Per Source $e^-$ & Norm.(70$^{\circ}$F)\\
\hline
Det. E Deposition (MeV/ptcl) ($\Delta/^{\circ}$F)  & $0.34(40)\times10^{-6}$ & $39.8(67)\times10^{-6}$  \\ 
Det. Window $e^-$ Current ($\Delta/^{\circ}$F)  & $20.3(13)\times10^{-6}$ & $35.3(33)\times10^{-6}$  \\ 
Det. Window E Current(MeV/ptcl) ($\Delta/^{\circ}$F)  & $4.32(32)\times10^{-6}$ & $32.4(34)\times10^{-6}$  \\ 
Det./Source $e^-$ Current Ratio ($\Delta/^{\circ}$F)  & $29.5(24)\times10^{-6}$ & $41.5(48)\times10^{-6}$  \\ 
Det./Source E Current Ratio ($\Delta/^{\circ}$F) &  $24.9(25)\times10^{-6}$ & $33.7(48)\times10^{-6}$  \\ 
\hline 
\end{tabular} 
\end{table}
\normalsize

Further analysis of climate as a cause of the observed oscillations can be carried out by examining the phase of the $^{36}$Cl data taken at OSURR, and comparing it to the phases of temperature, atmospheric pressure, and relative humidity data reported for Columbus, Ohio. The dominant frequency in the power spectrum of $^{36}$Cl provided in Figure \ref{fig:spec} is at 1.06$\pm$0.09 yr$^{-1}$, and it has an associated phase of -48.8$\pm$3.2$^{\circ}$, which corresponds to approximately 17 February, with 1 January equal to 0$^{\circ}$. This is consistent with the results of frequency and phase analyses carried out by our group for the other experiments (BNL, PTB and that from Ref. \cite{ell90}) presented in Refs. \citep{jav10,stu11b}. It is clear from Fig.~\ref{fig:data} that the $^{36}$Cl phase lags the phase of $1/\mathrm{R}^{2}$, but this was also the case for the BNL \cite{alb86} and PTB \cite{sie98} examined in Ref. \cite{jen09b}, in which we showed that the phases of the BNL and PTB data were similar for the two year period in which the experiments overlapped. One can also see from Fig.~\ref{fig:data} that the phase lag is not perfectly consistent across the six year period reported, a point to which we will return shortly. Thus, the reported phase of the $^{36}$Cl data, at -48.8$\pm$3.2$^{\circ}$, is an average, and gives no representation of phase stability.

We next turn to the comparison of the $^{36}$Cl data phase to the phases of the ambient temperature, pressure, and relative humidity in Columbus, Ohio, as reported by the National Climactic Data Center (NCDC) which is part of the National Oceanic and Atmospheric Administration (NOAA). The phases of ambient temperature, atmospheric pressure, and relative humidity were -16.7$^{\circ}$, -129$^{\circ}$, and -143.6$^{\circ}$ respectively. We should be careful to note here that the phases we are reporting for the climate data were obtained by sampling the climate parameters at the same sampling rate and dates as the $^{36}$Cl data (i.e., the data reported for the days which correspond to the days of the BC-4 calibrations). We should also note that the determined phases for the environmental variables are also averages, as for the $^{36}$Cl data, and do not give any indication of phase stability. In this case, as with the analysis by our group presented in Ref. \cite{jav10} we see that the phases of the weather variables or their reciprocals do not appear to match the phase of the measured $^{36}$Cl data.

Although the phase of the $^{36}$Cl data does not match that of $1/R^2$ exactly, this question was addressed by our group in Ref. \cite{stu11b}, where the difference is attributed to an inhomogeneous solar core. This inhomogeneity then results in a non-uniform solar $\nu_e$ production between the northern and southern solar hemispheres. Earth's maximum exposure to the southern hemisphere of the Sun is $\sim$8 March due to the tilt of the solar axis with respect to the plane of the Earth's orbit. Thus, the combination of the closer proximity at perihelion, and the differing exposures to more active regions of the Sun, could explain the phase lag. This phase lag is, in fact, very close to what is observed in data from two of the major neutrino detectors. Analysis of data from Super-Kamiokande-I in Japan \cite{smy04,hos06} provides a peak in the solar neutrino flux at perihelion plus 13$\pm$17 days \cite{smy04,hos06}, and places a 68\% confidence range for the actual peak that covers a range that includes the first week of February. Analysis of data from the Sudbury Neutrino Observatory (SNO) \cite{aha05,ran07} shows the best fit for peak solar neutrino flux at perihelion plus 40 days ($\sim$12 February) \cite{ran07}, which is within 2$\sigma$ ($\sigma\approx$3.25 days, see above) of the calculated phase of the OSURR data, or about 17 February.

Moreover, as described in Refs. \cite{jav10, stu11b}, the actual observed phase, if any, in nuclear decay data will be dependent on the sensitivity of the decaying isotopes to the external effect, and whether or not the isotope(s) being measured are parents or daughters. Clearly, if the isotope being measured is a second or third generation progeny of a radioactive parent, any perturbations to the decay rate of the parent will have to work their way down the decay chain to the daughter, grand-daughter, etc. Evidence of this fact was reported in the analysis of the BNL data \cite{alb86} in Refs. \cite{jav10,stu11b}. The $^{36}$Cl data and the $^{32}$Si (in which case, the $^{32}$P daughter was the nuclide measured) did not exhibit the same phases, even though they were measured in alternating counts, where each was counted for thirty minutes respectively, ten times during a ten hour run. Another example of differing isotopic sensitivities was noted by \citet{ell90}, where the $^{56}$Mn data collected on a sodium iodide detector exhibited an annual oscillation, but the $^{137}$Cs calibration source measurements taken on the same detector on the same days did not.

Based on the considerations in the preceding paragraphs, we have shown that temperature variations are not a likely explanation for the observed periodicity. However, to be thorough we can examine the possibility that any of the materials of the BC-4 housing, the source, or the source holder may have expanded or contracted due to thermal expansion. The materials which comprise the source tray and detector are steel (which has a linear expansion coefficient of 7.3$\times$10$^{-6}$in$\cdot$in$^{-1}\cdot ^{\circ}$F$^{-1}$, 13$\times$10$^{-6}$m$\cdot$m$^{-1}\cdot$K$^{-1}$) and aluminum (which has a linear expansion coefficient of 12.3$\times$10$^{-6}$in$\cdot$in$^{-1}\cdot ^{\circ}$F$^{-1}$, 22.2$\times$10$^{-6}$m$\cdot$m$^{-1}\cdot$K$^{-1}$). Since the aluminum thermal expansion coefficient is larger, we can use it as a bounding value and assume that any expansion in steel will be smaller. 

Examining the simplest conservative case, then, we assume that the steel tracks that hold the source tray expand such that the source-detector distance was changed. We will examine the system in three dimensions, with the $xy$-plane being the source holder and the $z$-axis running from the origin at the center of the source up through the detector. Examining a change of the $z$-axis first, and using the (larger) aluminum expansion coefficient, we find that for an initial source-detector distance of 2 mm and a 10$^{\circ}$C ($\sim$18$^{\circ}$F) change in temperature, we would realize only a 0.0004 mm change in the source-detector distance. 

Based on this analysis, we find that the expected material expansion in the $z$-direction could not lead to a count rate change similar to those observed, particularly since the expectation would be a linear dependence, and not an $r^2$, since we have what amounts to a 2$\pi$ geometry.  
Furthermore, we can also examine the possible shifts that would occur in the $xy$-plane as a result of thermal expansion. If we assume similar changes to the tray width and length that we expect on the $z$-axis, we can conclude that the expected changes would be similarly inconsequential. The result would be a shift on the order of $\sim$0.0202 mm  of what amounts to a small active source region ($\sim$ 0.4 cm$^{2}$ combined area) under what is, as we described previously, essentially a 2$\pi$ geometry of the pancake G-M detector, which has an active window of $\sim$15.5 cm$^{2}$. This too will not have enough impact to cause the observed variation in count rates.

\begin{figure*}[ht]
\includegraphics[scale=0.90]{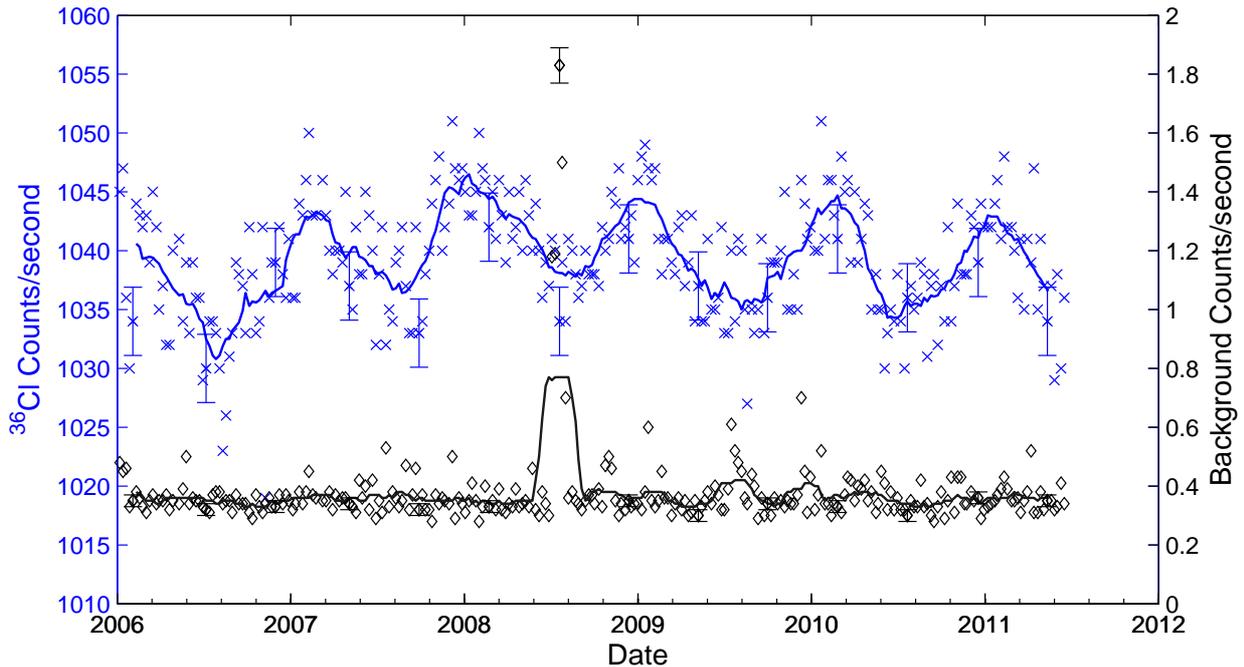}
\caption{Comparison of the measured $^{36}$Cl and background measurements (in counts/second) on the OSURR detector from January 2006 to June 2011. Representative error bars shown are standard Poisson ($\sqrt{N}$) uncertainties for the respective measurements. The large increase in the background countrate in 2008 was associated with the reactor control rods being temporarily stored near the detector system during a routine maintenance outage. \label{fig:bkdandcts}}
\end{figure*}

\begin{figure}[h]
\includegraphics[width=\columnwidth]{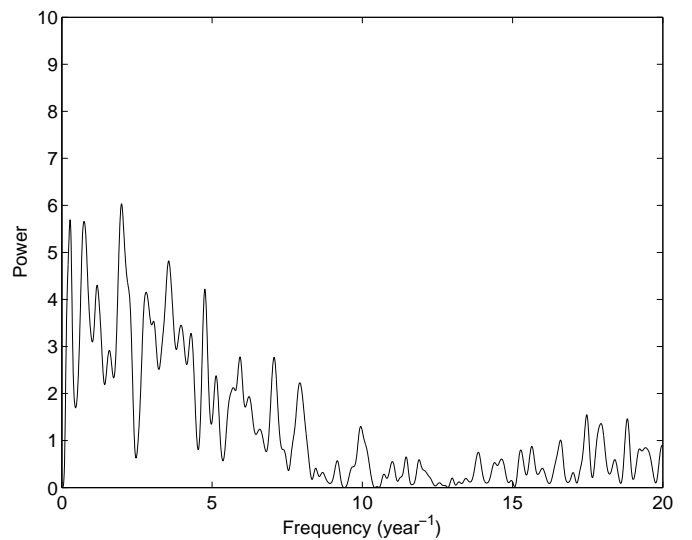}
\caption{Frequency spectrum for the OSURR measured background data from January 2006 through June 2011. \label{fig:bkg}}
\end{figure}

Having addressed the likely suspects for systematic and climactic influences, there remains the question of counts due to background. While the detector itself is well shielded, there will still be background counts included in the 2 minute gross count measurements. As noted above, 10 minute background counts were taken each week along with the $^{36}$Cl calibration count. The average of the weekly background counts for the period January 2006-June 2011 is 0.37$\pm$0.14 counts/s. When compared to the 1038$\pm$3 counts/sec for the $^{36}$Cl source, it is evident that changes in the background were not likely to be responsible for the decay rate variations. A point by point comparison of the $^{36}$Cl counts/sec to the measured background counts/sec is presented in Fig.~\ref{fig:bkdandcts}. As can be seen from this figure, while the annual periodicity is clearly evident in the $^{36}$Cl data, there is no similar periodicity evident in the background. This is verified by a power spectrum analysis similar to the one performed on the $^{36}$Cl data, which is presented in Fig.~\ref{fig:bkg}. No similar frequency structure is evident in these background data.

The measured background counts are dependent on what was happening in the laboratory area surrounding the detector at the time each count was taken, such as the reactor power at the time of the background count, and what radioactive materials happen to be located near the detector. For instance, the obvious spike in the background counts during the summer of 2008 was the result of the storage of the stainless steel reactor control rods (which are radioactive as a result of neutron-activation) nearby during a routine periodic reactor maintenance outage. Note from Fig.~\ref{fig:bkdandcts} that there was no discernible increase in the $^{36}$Cl count rate during that period; in fact, this increase in background occurred in the middle of the summer decrease in the $^{36}$Cl counts.  Providing additional support is the analysis of Ref. \cite{jen10}, which includes a rigorous examination of background effects in all types of gas detectors (specifically, the BNL and PTB detector systems). It was determined that in all cases, the dose rates to the detector from known backgrounds (such as cosmic radiation) were too low to have accounted for the oscillations in the BNL or PTB data, and by extension we can assume the same applies here.

There are some key conclusions that can be drawn from this analysis of these background data. First, changes in background could not have been the cause of the oscillations evident in the measured $^{36}$Cl decay rate during the recording period. The background data do not contain the same periodicity as the $^{36}$Cl standard measurements as seen in Figs. \ref{fig:spec}, \ref{fig:bkdandcts} and \ref{fig:bkg}. Moreover, even a $\sim$480\% increase in the background count did not appear to discernibly affect the measured $^{36}$Cl decay rate, with the elevated background still only comprising $\sim$0.18\% of the total counts (1034 counts/sec for the $^{36}$Cl, 1.83 cts/sec for background). Second, the absence of periodicities in the background similar to those shown the $^{36}$Cl data address two of our specific questions. One, the stability in the background data speak to the relative stability of the counting systems with respect to the influence of the environment on the detector or source. Two, the observed fluctuations are not the result of transients in the detector electronics, since no similar fluctuations appear in the background counts which were taken immediately before or after the $^{36}$Cl counts. We can now reasonably conclude that whatever is causing the count rates is not likely to be in the detector, or the environment surrounding the detector.

\section{Conclusions}

Having eliminated all of the possible known systematic causes of the fluctuations, we are left with the possibility that the measured decay rate changes are caused by an influence not of terrestrial origin, such as the Sun as suggested in Refs. \cite{jen09a,jen09b,fis09,stu10a,jav10,stu10b,stu11a,stu11b}. There is now apparent corroboration for the annual periodicity in the $^{36}$Cl data from the BNL experiment in Refs. \cite{alb86,jen09b}. While the sub-annual periodicities present in the BNL and PTB data sets as shown in Refs. \cite{stu10a,stu10b,stu11a} do not appear in this OSURR data, this is not surprising based on the statistics of the relatively small count rates (fractional error $\sim$0.3\%). In contrast, the $^{36}$Cl data of the BNL experiment did show additional periodicities \citep{alb86,stu10a,stu11a}, but the statistics of those counts were much better, with each data point consisting of $\sim$1.2$\times$10$^{6}$ counts (fractional error $\sim$0.03\%). However, the annual oscillation is clearly evident in the OSURR data, and thus lends support to the conclusions of Refs. \cite{jen09a,jen09b,fis09,stu10a,stu10b,stu11a,stu11b,ste11}, while providing additional evidence of the annual effect in $^{36}$Cl from more recently measured decay data.

There is an obvious need for more experiments examining a wide variety of detector technologies and a large array of isotopes utilizing appropriate controls, including directly measuring temperature, pressure and humidity effects on detector systems in environmental chambers. There is also a need to examine records of experiments and calibrations such as this one, where data are available that could be analyzed for the presence of unexplained oscillations in what should be randomly distributed data. A concerted effort will be required by many to address the questions that remain about the causes of these fluctuations, including the development of a physical model. Hopefully this will also lead to more accurate determinations of half-lives for the $\sim$3300 known radioactive isotopes, and a better understanding of the physical world.

\section*{Acknowledgments}
The authors wish to thank Bob Leino of Thermo Fisher Scientific for providing the technical information and drawings for the Eberline BC-4,  and also A. Treacher and H. Miser for their helpful assistance. The work of EF is supported in part by U.S. DOE contract No. DE-AC02-76ER071428, and the work of PAS was supported in part by the NSF through Grant AST-06072572.





\bibliographystyle{model1a-num-names}

\begin{thebibliography}{48}
\expandafter\ifx\csname natexlab\endcsname\relax\def\natexlab#1{#1}\fi
\providecommand{\bibinfo}[2]{#2}
\ifx\xfnm\relax \def\xfnm[#1]{\unskip,\space#1}\fi
\bibitem[{Jenkins and Fischbach(2009)}]{jen09a}
\bibinfo{author}{J.~H. Jenkins}, \bibinfo{author}{E.~Fischbach},
  \bibinfo{journal}{Astroparticle Physics} \bibinfo{volume}{31}
  (\bibinfo{year}{2009}) \bibinfo{pages}{407}.
\bibitem[{Jenkins et~al.(2009)Jenkins, Fischbach, Buncher, Gruenwald, Krause,
  and Mattes}]{jen09b}
\bibinfo{author}{J.~H. Jenkins}, \bibinfo{author}{et~al.},
    \bibinfo{journal}{Astroparticle Physics} \bibinfo{volume}{32}
  (\bibinfo{year}{2009}) \bibinfo{pages}{42}.
\bibitem[{Fischbach et~al.(2009)Fischbach, Buncher, Gruenwald, Jenkins, Krause,
  Mattes, and Newport}]{fis09}
\bibinfo{author}{E.~Fischbach}, \bibinfo{author}{et~al.}, \bibinfo{journal}{Space Science Reviews}
  \bibinfo{volume}{145} (\bibinfo{year}{2009}) \bibinfo{pages}{285}.
\bibitem[{Alburger et~al.(1986)Alburger, Harbottle, and Norton}]{alb86}
\bibinfo{author}{D.~E. Alburger}, \bibinfo{author}{G.~Harbottle},
  \bibinfo{author}{E.~F. Norton}, \bibinfo{journal}{Earth and Planetary Science
  Letters} \bibinfo{volume}{78} (\bibinfo{year}{1986}) \bibinfo{pages}{168}.
\bibitem[{Siegert et~al.(1998)Siegert, Schrader, and Sch\"{o}tzig}]{sie98}
\bibinfo{author}{H.~Siegert}, \bibinfo{author}{H.~Schrader},
  \bibinfo{author}{U.~Sch\"{o}tzig}, \bibinfo{journal}{Applied Radiation and
  Isotopes} \bibinfo{volume}{49} (\bibinfo{year}{1998}) \bibinfo{pages}{1397}.
\bibitem[{Javorsek~II et~al.(2010)Javorsek~II, Sturrock, Lasenby, Lasenby,
  Buncher, Fischbach, Gruenwald, Hoft, Horan, Jenkins, Kerford, Lee, Longman,
  Mattes, Morreale, Morris, Mudry, Newport, O'Keefe, Petrelli, Silver, Stewart,
  and Terry}]{jav10}
\bibinfo{author}{D.~Javorsek~II}, \bibinfo{author}{et~al.},
 \bibinfo{journal}{Astroparticle Physics}
  \bibinfo{volume}{34} (\bibinfo{year}{2010}) \bibinfo{pages}{173}.
\bibitem[{Sturrock et~al.(2010{\natexlab{a}})Sturrock, Buncher, Fischbach,
  Gruenwald, Javorsek~II, Jenkins, Lee, Mattes, and Newport}]{stu10a}
\bibinfo{author}{P.~A. Sturrock}, \bibinfo{author}{et~al.}, \bibinfo{journal}{Astroparticle Physics}
  \bibinfo{volume}{34} (\bibinfo{year}{2010}{\natexlab{a}})
  \bibinfo{pages}{121}.
\bibitem[{Sturrock et~al.(2010{\natexlab{b}})Sturrock, Buncher, Fischbach,
  Gruenwald, Javorsek, Jenkins, Lee, Mattes, and Newport}]{stu10b}
\bibinfo{author}{P.~A. Sturrock}, \bibinfo{author}{et~al.}, \bibinfo{journal}{Solar Physics}
  \bibinfo{volume}{267} (\bibinfo{year}{2010}{\natexlab{b}})
  \bibinfo{pages}{251}.
\bibitem[{Sturrock et~al.(1999)Sturrock, Scargle, Walther, and
  Wheatland}]{stu99}
\bibinfo{author}{P.~A. Sturrock}, \bibinfo{author}{et~al.},
  \bibinfo{journal}{The Astrophysical Journal Letters} \bibinfo{volume}{523}
  (\bibinfo{year}{1999}) \bibinfo{pages}{L177}.
\bibitem[{Sturrock(2008)}]{stu08}
\bibinfo{author}{P.~A. Sturrock}, \bibinfo{journal}{The Astrophysical Journal
  Letters} \bibinfo{volume}{688} (\bibinfo{year}{2008}) \bibinfo{pages}{L53}.
\bibitem[{Sturrock(2009)}]{stu09}
\bibinfo{author}{P.~A. Sturrock}, \bibinfo{journal}{Solar Physics}
  \bibinfo{volume}{254} (\bibinfo{year}{2009}) \bibinfo{pages}{227}.
\bibitem[{Sturrock et~al.(2011)Sturrock, Fischbach, and Jenkins}]{stu11a}
\bibinfo{author}{P.~A. Sturrock}, \bibinfo{author}{E.~Fischbach},
  \bibinfo{author}{J.~H. Jenkins}, \bibinfo{journal}{Solar Physics}
  \bibinfo{volume}{272} (\bibinfo{year}{2011}) \bibinfo{pages}{1--10}.
\bibitem[{Rieger et~al.(1984)Rieger, Share, Forrest, Kanbach, Reppin, and
  Chupp}]{rie84}
\bibinfo{author}{E.~Rieger}, \bibinfo{author}{et~al.},
  \bibinfo{journal}{Nature} \bibinfo{volume}{312} (\bibinfo{year}{1984})
  \bibinfo{pages}{623}.
\bibitem[{Sturrock et~al.(2011)Sturrock, Buncher, Fischbach, Javorsek~II,
  Jenkins, and Mattes}]{stu11b}
\bibinfo{author}{P.~Sturrock}, \bibinfo{author}{et~al.},
  \bibinfo{journal}{The Astrophysical Journal} \bibinfo{volume}{737}
  (\bibinfo{year}{2011}) \bibinfo{pages}{65}.
\bibitem[{Norman et~al.(2009)Norman, Browne, Shugart, Joshi, and
  Firestone}]{nor09}
\bibinfo{author}{E.~B. Norman}, \bibinfo{author}{et~al.}, \bibinfo{journal}{Astroparticle Physics}
  \bibinfo{volume}{31} (\bibinfo{year}{2009}) \bibinfo{pages}{135}.
\bibitem[{Silverman and Strange(2009)}]{sil09}
\bibinfo{author}{M.~P. Silverman}, \bibinfo{author}{W.~Strange},
  \bibinfo{journal}{EPL (Europhysics Letters)} \bibinfo{volume}{87}
  (\bibinfo{year}{2009}) \bibinfo{pages}{32001}.
\bibitem[{Cooper(2009)}]{coo09}
\bibinfo{author}{P.~S. Cooper}, \bibinfo{journal}{Astroparticle Physics}
  \bibinfo{volume}{31} (\bibinfo{year}{2009}) \bibinfo{pages}{267}.
\bibitem[{O'Keefe et~al.(2012)O'Keefe, Morreale, Fischbach, Javorsek~II,
  Jenkins, Lee, Morris, and Sturrock}]{oke11}
\bibinfo{author}{D.~O'Keefe}, \bibinfo{author}{et~al.},
  \bibinfo{journal}{Astroparticle Physics}  (\bibinfo{year}{2012})
  \bibinfo{pages}{Under Review}.
\bibitem[{Jenkins et~al.(2010)Jenkins, Mundy, and Fischbach}]{jen10}
\bibinfo{author}{J.~H. Jenkins}, \bibinfo{author}{D.~W. Mundy},
  \bibinfo{author}{E.~Fischbach}, \bibinfo{journal}{Nuclear Instruments and
  Methods in Physics Research Section A: Accelerators, Spectrometers, Detectors
  and Associated Equipment} \bibinfo{volume}{620} (\bibinfo{year}{2010})
  \bibinfo{pages}{332}.
\bibitem[{Center(2011)}]{nndc}
\bibinfo{author}{N.~N.~D. Center}, \bibinfo{title}{Information extracted from
  the chart of nuclides database}, \bibinfo{year}{2011}.
\bibitem[{Parkhomov(2005)}]{par05}
\bibinfo{author}{A.~G. Parkhomov}, \bibinfo{journal}{International Journal of
  Pure and Applied Physics} \bibinfo{volume}{1} (\bibinfo{year}{2005})
  \bibinfo{pages}{119}.
\bibitem[{Parkhomov(2010{\natexlab{a}})}]{par10a}
\bibinfo{author}{A.~G. Parkhomov}, \bibinfo{journal}{ArXiv}
  \bibinfo{volume}{arXiv:1012.4174v1 [physics.gen-ph]}
  (\bibinfo{year}{2010}{\natexlab{a}}) \bibinfo{pages}{1}.
\bibitem[{Parkhomov(2010{\natexlab{b}})}]{par10b}
\bibinfo{author}{A.~G. Parkhomov}, \bibinfo{journal}{ArXiv}
  \bibinfo{volume}{arXiv:1004.1761v1 [physics.gen-ph]}
  (\bibinfo{year}{2010}{\natexlab{b}}).
\bibitem[{Ellis(1990)}]{ell90}
\bibinfo{author}{K.~J. Ellis}, \bibinfo{journal}{Physics in Medicine and
  Biology} \bibinfo{volume}{35} (\bibinfo{year}{1990}) \bibinfo{pages}{1079}.
\bibitem[{Steinitz et~al.(2011)Steinitz, Piatibratova, and Kotlarsky}]{ste11}
\bibinfo{author}{G.~Steinitz}, \bibinfo{author}{O.~Piatibratova},
  \bibinfo{author}{P.~Kotlarsky}, \bibinfo{journal}{Journal of Environmental
  Radioactivity} \bibinfo{volume}{102} (\bibinfo{year}{2011})
  \bibinfo{pages}{749}.
\bibitem[{Sturrock et~al.(2012)Sturrock, Parkhomov, Fischbach, and
  Jenkins}]{stu12}
\bibinfo{author}{P.~A. Sturrock}, \bibinfo{author}{et~al.},
  \bibinfo{journal}{Astroparticle Physics}  (\bibinfo{year}{2012})
  \bibinfo{pages}{In press}.
\bibitem[{Lindstrom et~al.(2010)Lindstrom, Fischbach, Buncher, Greene, Jenkins,
  Krause, Mattes, and Yue}]{lin10}
\bibinfo{author}{R.~M. Lindstrom}, \bibinfo{author}{et~al.},
  \bibinfo{journal}{Nuclear Instruments and Methods in Physics Research Section
  A: Accelerators, Spectrometers, Detectors and Associated Equipment}
  \bibinfo{volume}{622} (\bibinfo{year}{2010}) \bibinfo{pages}{93--96}.
  \bibinfo{note}{Doi: DOI: 10.1016/j.nima.2010.06.270}.
\bibitem[{Lindstrom et~al.(2011)Lindstrom, Fischbach, Buncher, Jenkins, and
  Yue}]{lin11}
\bibinfo{author}{R.~M. Lindstrom}, \bibinfo{author}{et~al.}, \bibinfo{journal}{Nuclear Instruments and Methods
  in Physics Research Section A: Accelerators, Spectrometers, Detectors and
  Associated Equipment} \bibinfo{volume}{659} (\bibinfo{year}{2011})
  \bibinfo{pages}{269--271}.
\bibitem[{de~Meijer et~al.(2011)de~Meijer, Blaauw, and Smit}]{dem11}
\bibinfo{author}{R.~J. de~Meijer}, \bibinfo{author}{M.~Blaauw},
  \bibinfo{author}{F.~D. Smit}, \bibinfo{journal}{Applied Radiation and
  Isotopes} \bibinfo{volume}{69} (\bibinfo{year}{2011})
  \bibinfo{pages}{320--326}.
\bibitem[{ebe(1972)}]{ebe72}
\bibinfo{title}{Technical Manual for Beta Counter Model BC-4},
  \bibinfo{publisher}{Eberline Instrument Corporation}, \bibinfo{address}{Santa
  Fe}, \bibinfo{year}{1972}.
\bibitem[{Endt and Firestone(1999)}]{end99}
\bibinfo{author}{P.~M. Endt}, \bibinfo{author}{R.~B. Firestone}
  (\bibinfo{year}{1999}).
\bibitem[{Scargle(1982)}]{sca82}
\bibinfo{author}{J.~D. Scargle}, \bibinfo{journal}{Astrophysical Journal}
  \bibinfo{volume}{263} (\bibinfo{year}{1982}) \bibinfo{pages}{835}.
\bibitem[{Fujioka et~al.(1951)Fujioka, Kita, and Minakawa}]{fuj51}
\bibinfo{author}{G.~Fujioka}, \bibinfo{author}{I.~Kita},
  \bibinfo{author}{O.~Minakawa}, \bibinfo{journal}{Journal of the Physical
  Society of Japan} \bibinfo{volume}{6} (\bibinfo{year}{1951})
  \bibinfo{pages}{103}.
\bibitem[{Haq(1983)}]{haq83}
\bibinfo{author}{F.~U. Haq}, \bibinfo{journal}{Journal of Physics E: Scientific
  Instruments} \bibinfo{volume}{16} (\bibinfo{year}{1983})
  \bibinfo{pages}{724}.
\bibitem[{Khristov and Kirov(1960)}]{khr60}
\bibinfo{author}{L.~G. Khristov}, \bibinfo{author}{K.~I. Kirov},
  \bibinfo{journal}{Comptes Rendus de l'Academie Bulgare des Sciences}
  \bibinfo{volume}{13} (\bibinfo{year}{1960}) \bibinfo{pages}{399}.
\bibitem[{Kimura(1950)}]{kim50}
\bibinfo{author}{M.~Kimura}, \bibinfo{journal}{Physical Review}
  \bibinfo{volume}{80} (\bibinfo{year}{1950}) \bibinfo{pages}{761}.
\bibitem[{Kimura(1951)}]{kim51}
\bibinfo{author}{M.~Kimura}, \bibinfo{journal}{Journal of the Physical Society
  of Japan} \bibinfo{volume}{6} (\bibinfo{year}{1951}) \bibinfo{pages}{141}.
\bibitem[{Liebson and Friedman(1948)}]{lie48}
\bibinfo{author}{S.~H. Liebson}, \bibinfo{author}{H.~Friedman},
  \bibinfo{journal}{Review of Scientific Instruments} \bibinfo{volume}{19}
  (\bibinfo{year}{1948}) \bibinfo{pages}{303}.
\bibitem[{Seidl(1952)}]{sei52a}
\bibinfo{author}{R.~Seidl}, \bibinfo{journal}{Czechoslovak Journal of Physics}
  \bibinfo{volume}{1} (\bibinfo{year}{1952}) \bibinfo{pages}{160}.
\bibitem[{Semkow et~al.(2009)Semkow, Haines, Beach, Kilpatrick, Khan, and
  O'Brien}]{sem09}
\bibinfo{author}{T.~M. Semkow}, \bibinfo{author}{et~al.},
  \bibinfo{journal}{Physics Letters B} \bibinfo{volume}{675}
  (\bibinfo{year}{2009}) \bibinfo{pages}{415}.
\bibitem[{Davis(1992)}]{dav92}
\bibinfo{author}{R.~S. Davis}, \bibinfo{journal}{Metrologia}
  \bibinfo{volume}{29} (\bibinfo{year}{1992}) \bibinfo{pages}{67}.
\bibitem[{Picard et~al.(2008)Picard, Davis, Glaser, and Fujii}]{pic08}
\bibinfo{author}{A.~Picard}, \bibinfo{author}{et~al.},
  \bibinfo{journal}{Metrologia} \bibinfo{volume}{45} (\bibinfo{year}{2008})
  \bibinfo{pages}{149}.
\bibitem[{Hendricks et~al.(2008)Hendricks, McKinney, Fensin, James, Johns,
  Durkee, Finch, Pelowitz, Waters, Johnson, and Gallmeier}]{mcnp}
\bibinfo{author}{J.~S. Hendricks}, \bibinfo{author}{et~al.}, \bibinfo{title}{MCNPX 2.6.0 Extensions},
  \bibinfo{type}{Technical Report}, LANL, \bibinfo{year}{2008}.
\bibitem[{Found(1920)}]{fou20}
\bibinfo{author}{C.~G. Found}, \bibinfo{journal}{Physical Review}
  \bibinfo{volume}{16} (\bibinfo{year}{1920}) \bibinfo{pages}{41}.
\bibitem[{Smy et~al.(2004)Smy, Ashie, Fukuda, Fukuda, Ishihara, Itow, Koshio,
  Minamino, Miura, Moriyama, Nakahata, Namba, Nambu, Obayashi, Sakurai,
  Shiozawa, Suzuki, Takeuchi, Takeuchi, Yamada, Ishitsuka, Kajita, Kaneyuki,
  Nakayama, Okada, Ooyabu, Saji, Desai, Earl, Kearns, Messier, Stone, Sulak,
  Walter, Wang, Goldhaber, Barszczak, Casper, Gajewski, Kropp, Mine, Liu,
  Sobel, Vagins, Gago, Ganezer, Hill, Keig, Kim, Lim, Ellsworth, Tasaka,
  Kibayashi, Learned, Matsuno, Takemori, Hayato, Ichikawa, Ishii, Kameda,
  Kobayashi, Maruyama, Nakamura, Nitta, Oyama, Sakuda, Totsuka, Yoshida,
  Iwashita, Suzuki, Inagaki, Kato, Nakaya, Nishikawa, Haines, Dazeley,
  Hatakeyama, Svoboda, Blaufuss, Goodman, Guillian, Sullivan, Turcan,
  Scholberg, Habig, Ackermann, Jung, Kato, Kobayashi, Martens, Malek, Mauger,
  McGrew, Sharkey, Viren, Yanagisawa, Toshito, Mitsuda, Miyano, Shibata
  et~al.}]{smy04}
\bibinfo{author}{M.~B. Smy}, et~al.,
  \bibinfo{journal}{Physical Review D} \bibinfo{volume}{69}
  (\bibinfo{year}{2004}) \bibinfo{pages}{011104}. \bibinfo{note}{PRD
  (Super-Kamiokande Collaboration)}.
\bibitem[{Hosaka et~al.(2006)Hosaka, Ishihara, Kameda, Koshio, Minamino,
  Mitsuda, Miura, Moriyama, Nakahata, Namba, Obayashi, Sakurai, Sarrat,
  Shiozawa, Suzuki, Takeuchi, Yamada, Higuchi, Ishitsuka, Kajita, Kaneyuki,
  Mitsuka, Nakayama, Nishino, Okada, Okumura, Saji, Takenaga, Clark, Desai,
  Kearns, Likhoded, Stone, Sulak, Wang, Goldhaber, Casper, Cravens, Kropp, Liu,
  Mine, Smy, Sobel, Sterner, Vagins, Ganezer, Hill, Keig, Jang, Kim, Lim,
  Scholberg, Walter, Ellsworth, Tasaka, Guillian, Kibayashi, Learned, Matsuno,
  Messier, Hayato, Ichikawa, Ishida, Ishii, Iwashita, Kobayashi, Nakadaira,
  Nakamura, Nitta, Oyama, Totsuka, Suzuki, Hasegawa, Kato, Maesaka, Nakaya,
  Nishikawa, Sasaki, Sato, Yamamoto, Yokoyama, Haines, Dazeley, Kim, Lee,
  Hatakeyama, Svoboda, Blaufuss, Goodman, Sullivan, Turcan, Cooley, Habig,
  Fukuda, Sato, Itow, Jung, Kato, Kobayashi, Malek et~al.}]{hos06}
\bibinfo{author}{J.~Hosaka}, et~al., \bibinfo{journal}{Physical Review D}
  \bibinfo{volume}{73} (\bibinfo{year}{2006}) \bibinfo{pages}{112001}.
  \bibinfo{note}{PRD}.
\bibitem[{Aharmim et~al.(2005)Aharmim, Ahmed, Anthony, Beier, Bellerive,
  Bergevin, Biller, Boulay, Bowler, Chan, Chen, Chen, Cleveland, Costin, Cox,
  Currat, Dai, Deng, Detwiler, Doe, Dosanjh, Doucas, Duba, Duncan, Dunford,
  Dunmore, Earle, Elliott, Evans, Ewan, Farine, Fergani, Fleurot, Formaggio,
  Frati, Fulsom, Gagnon, Goon, Graham, Hahn, Hallin, Hallman, Handler,
  Hargrove, Harvey, Hazama, Heeger, Heelan, Heintzelman, Heise, Helmer,
  Hemingway, Hime, Howe, Huang, Inrig, Jagam, Jelley, Klein, Kormos, Kos,
  Krüger, Kraus, Krauss, Krumins, Kutter, Kyba, Labranche, Lange, Law, Lawson,
  Lesko, Leslie, Levine, Loach, Luoma, MacLellan, Majerus, Maneira, Marino,
  McCauley, McDonald, McGee, Mifflin, Miknaitis, Nickel, Noble, Norman, Oblath,
  Okada, O'Keeffe, Ollerhead, Gann, Orrell, Oser, Ouvarova, Peeters, Poon, Pun,
  Rielage et~al.}]{aha05}
\bibinfo{author}{B.~Aharmim}, et~al., \bibinfo{journal}{Physical Review D}
  \bibinfo{volume}{72} (\bibinfo{year}{2005}) \bibinfo{pages}{052010}.
  \bibinfo{note}{PRD}.
\bibitem[{Ranucci and Rovere(2007)}]{ran07}
\bibinfo{author}{G.~Ranucci}, \bibinfo{author}{M.~Rovere},
  \bibinfo{journal}{Physical Review D} \bibinfo{volume}{75}
  (\bibinfo{year}{2007}) \bibinfo{pages}{013010}. \bibinfo{note}{PRD}.

\end{thebibliography}



\end{document}